\documentclass[10pt,twocolumn]{article}

\usepackage[margin=0.75in]{geometry}
\usepackage{times}
\usepackage{graphicx}
\usepackage{amsmath,amssymb}
\usepackage{booktabs}
\usepackage{siunitx}
\usepackage{hyperref}
\usepackage{caption}
\usepackage{subcaption}

\hypersetup{colorlinks=true, linkcolor=blue, citecolor=blue, urlcolor=blue}

\title{\bf Fisher-Information--Driven Adaptive Acquisition for Photon-Efficient FLIM:\\
A Dual-Implementation Framework for TCSPC and Programmable Time-Gating}

\author{J.\ Sumaya-Martinez$^{1,*}$, Eugenio Torres-Garcia$^{2}$\\
\small $^{1}$Laboratorio de Bio-fotónica, Facultad de Ciencias, Universidad Autónoma del Estado de México, Toluca, 50000, México\\
\small $^{2}$Laboratorio de Dosimetría y Simulación Monte Carlo, Facultad de Medicina, Universidad Autónoma del Estado de México, Toluca, 50180, Mexico\\
\small ORCID: \href{https://orcid.org/0000-0002-7032-8824}{0000-0002-7032-8824} (J.S.-M.),
\href{https://orcid.org/0000-0001-8355-3927}{0000-0001-8355-3927} (E.T.-G.)\\
\small $^{*}$Corresponding author: \href{mailto:jsm@uaemex.mx}{jsm@uaemex.mx}}

\date{}

\begin{document}
\maketitle

\begin{abstract}
Fluorescence lifetime imaging microscopy (FLIM) provides quantitative biochemical contrast but is often photon-limited in living specimens, where increased excitation dose leads to phototoxicity and photobleaching. Fisher information (FI) is widely used to derive Cram\'er--Rao lower bounds (CRLB) and to formalize biochemical resolving power, yet it is rarely exploited to \emph{actively design} the acquisition under realistic constraints such as finite instrument response functions (IRFs) and background. Building on FI-based biochemical resolving power \cite{Esposito2013BRP} and FLIM-specific FI analyses that clarify the impact of IRF and photon statistics \cite{Trinh2021BRPFLIM}, we develop a nuisance-robust FI framework that turns FI from a diagnostic bound into an adaptive controller for temporal sampling.

We derive FI expressions for Poisson photon counting in time-domain FLIM for a bi-exponential decay model convolved with a measured IRF and additive background. To handle practical uncertainties, we compute an \emph{effective} FI for biochemical parameters by marginalizing nuisance parameters (background and IRF) via the Schur complement, consistent with realistic lifetime-precision studies \cite{Bouchet2019CRLB}. At each iteration, the next temporal sampling design (TCSPC binning or time-gate placement/width) is selected by maximizing a D-optimal utility (log-determinant of the effective FI) under an explicit photon (dose) budget. We provide two Biomedical Optics Express-ready implementations: (I) adaptive TCSPC re-binning between frames/iterations, and (II) programmable time-gating on camera/SPAD platforms. Simulations demonstrate substantial photon-efficiency gains and improved robustness compared to uniform sampling, with Monte Carlo errors approaching CRLB trends. The accompanying Overleaf package includes reproducible figures and a manuscript scaffold designed for straightforward replacement with phantom and live-cell datasets.
\end{abstract}

\section{Introduction}
Quantitative fluorescence lifetime imaging microscopy (FLIM) provides contrast linked to biochemical state (e.g., microenvironment, binding, or metabolism) while being less sensitive than intensity measurements to excitation and detection fluctuations. In living specimens, however, FLIM is frequently \emph{photon-limited}: improving parameter precision by increasing dwell time or excitation power can exacerbate photobleaching and phototoxicity. A central question in dose-limited imaging is therefore how to allocate limited photons to maximize information about biochemical parameters.

Fisher information theory offers a principled connection between photon statistics, instrument design, and achievable parameter precision through the Cram\'er--Rao lower bound (CRLB). Esposito \emph{et al.} introduced \emph{biochemical resolving power} using FI to characterize and optimize fluorescence detection across modalities, including lifetime and multi-dimensional detection \cite{Esposito2013BRP}. Trinh and Esposito extended this analysis to time-domain FLIM and clarified how the instrument response function (IRF) and photon statistics jointly limit biochemical resolvability \cite{Trinh2021BRPFLIM}. Complementary CRLB analyses further emphasize that IRF properties, background, and model mismatch can dominate lifetime precision in practice \cite{Bouchet2019CRLB}.

Despite these advances, FI is typically used \emph{offline} to diagnose limits for a fixed acquisition rather than to \emph{control} the acquisition itself. In practice, many FLIM systems offer control over temporal sampling: the binning of TCSPC histograms, the placement and width of time gates, and the allocation of dwell time across gates. These choices determine how efficiently detected photons constrain biochemical parameters, and the optimal design can vary across samples, lifetimes, and noise regimes. This motivates a closed-loop approach: use a low-dose \emph{scout} measurement to estimate parameters (and uncertainties), then adapt the temporal sampling to maximize expected information for the remaining dose budget.

\paragraph{Contributions.}
(i) A unified FI model for Poisson time-domain FLIM with IRF convolution and background; (ii) nuisance-robust effective FI for biochemical parameters via the Schur complement; (iii) an adaptive design algorithm that selects temporal sampling to maximize D-optimal utility; and (iv) two practical implementations---TCSPC re-binning and programmable time-gating---aligned with common biomedical optics workflows.

\begin{figure*}[t]
\centering
\includegraphics[width=0.95\linewidth]{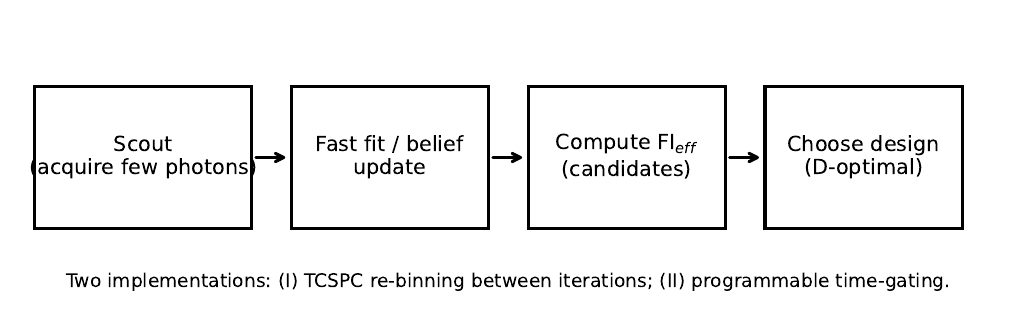}
\caption{FI-driven adaptive acquisition. A low-dose scout acquisition initializes parameter beliefs, which are used to compute the effective FI for candidate temporal samplings. The next sampling design is chosen by optimizing an information-theoretic utility under constraints, and the remaining photon budget is allocated accordingly.}
\label{fig:workflow}
\end{figure*}

\section{Methods}
\subsection{Observation model for time-domain FLIM}
We consider a pixel (or ROI) measured using $K$ temporal channels (bins or gates). Detected photon counts are modeled as independent Poisson variables,
\begin{equation}
y_k \sim \mathrm{Poisson}\!\left(\lambda_k(\theta,\phi; d)\right), \quad k=1,\dots,K,
\end{equation}
where $d$ denotes the temporal sampling \emph{design} (bin edges or gate parameters), $\theta$ are biochemical parameters of interest, and $\phi$ are nuisance parameters (background and IRF parameters). This photon-counting model and FI formulation are standard in microscopy \cite{Chao2016FItutorial}.

\subsection{Bi-exponential decay convolved with IRF}
A common biochemical FLIM model (e.g., metabolic NADH) is a bi-exponential decay,
\begin{equation}
f(t;\theta) = (1-a)\,e^{-t/\tau_f} + a\,e^{-t/\tau_b}, \quad \theta=\{a,\tau_f,\tau_b\},
\end{equation}
with bound fraction $a\in[0,1]$ and lifetimes $\tau_f,\tau_b$. The measured temporal response includes an IRF $h(t;\phi)$; the normalized convolved signal is
\begin{equation}
s(t;\theta,\phi)=\frac{(h * f)(t)}{\int_0^\infty (h * f)(u)\,du}.
\end{equation}
For a temporal bin (or gate) $\mathcal{B}_k(d)$, the expected count is
\begin{equation}
\lambda_k = N \int_{t\in \mathcal{B}_k(d)} s(t;\theta,\phi)\,dt + B_k(\phi),
\end{equation}
where $N$ is the expected signal photon budget and $B_k$ accounts for background photons integrated over the same window. Modeling finite IRF and background is essential for realistic precision limits \cite{Bouchet2019CRLB,Trinh2021BRPFLIM}.

\subsection{Fisher information for Poisson counts}
For independent Poisson channels, the Fisher information matrix (FIM) for parameters $\psi=[\theta,\phi]$ is
\begin{equation}
F_{ij}(\psi; d)=\sum_{k=1}^{K}\frac{1}{\lambda_k(\psi;d)}
\frac{\partial \lambda_k(\psi;d)}{\partial \psi_i}
\frac{\partial \lambda_k(\psi;d)}{\partial \psi_j}.
\label{eq:fim_poisson}
\end{equation}

\subsection{Nuisance-robust effective FI via Schur complement}
Partition the FIM into blocks,
\begin{equation}
F=\begin{bmatrix}
F_{\theta\theta} & F_{\theta\phi}\\
F_{\phi\theta} & F_{\phi\phi}
\end{bmatrix}.
\end{equation}
The nuisance-robust information about $\theta$ is quantified by the effective FIM,
\begin{equation}
F_{\mathrm{eff}} = F_{\theta\theta} - F_{\theta\phi}F_{\phi\phi}^{-1}F_{\phi\theta}.
\label{eq:schur}
\end{equation}
Optimizing $F_{\mathrm{eff}}$ yields designs robust to uncertainty in background and IRF \cite{Bouchet2019CRLB,Trinh2021BRPFLIM}.

\subsection{Design objective and constraints}
We select temporal sampling designs that maximize information about $\theta$ per photon under a budget. We use D-optimality,
\begin{equation}
U(d)=\log \det\!\left(F_{\mathrm{eff}}(\hat\psi; d)\right),
\end{equation}
where $\hat\psi$ is a current estimate or representative point from a belief distribution. Constraints include photon budget, acquisition time, and hardware limits (minimum gate width, number of gates).

\subsection{Adaptive acquisition: two-stage and iterative variants}
A practical two-stage scheme is:
\begin{enumerate}
\item \textbf{Scout:} Acquire a small fraction of the budget using a generic sampling (e.g., uniform bins) and estimate $\hat\psi$ by fast fitting or a Laplace approximation around a maximum likelihood estimate.
\item \textbf{FI-opt:} Choose a design from a candidate set $\mathcal{D}$ maximizing $U(d)$ and allocate the remaining budget to that design.
\end{enumerate}
Multi-iteration updates (repeating scout/optimize) can further improve performance when time allows.

\subsection{Standard equipment models and acquisition settings}
To make the framework directly actionable, we describe two widely used hardware configurations that map naturally onto our two implementations: a confocal TCSPC FLIM platform and a programmable time-gated FLIM platform.

\paragraph{(I) Confocal TCSPC FLIM (module-based).}
A representative configuration is a laser-scanning confocal microscope equipped with a pulsed excitation source (typical repetition rate 20--80~MHz) and a single-photon detector, coupled to a commercial TCSPC module such as the Becker \& Hickl \emph{SPC-150N} series \cite{BH_SPC150N_pdf,BH_SPC150N_page} or a multichannel TCSPC/time-tagging unit such as the PicoQuant \emph{HydraHarp 500} \cite{PQ_HydraHarp500_page,PQ_HydraHarp500_datasheet}. For our adaptive TCSPC implementation, the scout acquisition is collected with uniform binning; parameters are estimated; and an optimized (typically non-uniform) binning is selected for the remaining frames/iterations. The IRF is measured using a fast scatterer/SHG reference and included as a nuisance parameter in $F_{\mathrm{eff}}$.

\paragraph{(II) Programmable time-gated FLIM (camera/SPAD).}
A representative configuration is a camera-based gated FLIM attachment such as the Lambert Instruments \emph{LiFA-FLIM} platform \cite{Lambert_LiFAFLIM_page} mounted on a widefield microscope with pulsed excitation. Here the temporal design variables are the gate start times and widths. After a short scout acquisition using a small number of gates, the FI controller selects an optimized set of gates for the remaining dose budget. When hardware supports rapid updates, the design is adapted frame-to-frame; otherwise, it is adapted between short sequences.

\paragraph{Typical settings used in this manuscript.}
Table~\ref{tab:settings} lists practical parameter ranges used in simulations and suggested starting points for experiments.

\begin{table}[t]
\centering
\caption{Typical acquisition settings (starting points) for the two standard platforms.}
\label{tab:settings}
\begin{tabular}{@{}lll@{}}
\toprule
Parameter & TCSPC (confocal) & Time-gated (camera/SPAD) \\ \midrule
Time window & 0--12.5 ns & 0--12.5 ns \\
Channels & 16--256 bins & 4--16 gates \\
Scout budget & 5--10\% photons & 1--2 frames (few gates) \\
Optimized budget & 90--95\% photons & remaining frames \\
IRF handling & measured IRF; nuisance & measured IRF; nuisance \\
Background & constant or per-bin & constant or per-gate \\
\bottomrule
\end{tabular}
\end{table}

\section{Results}
We report simulations using a bi-exponential decay with finite IRF and background, motivated by FI-based FLIM studies \cite{Trinh2021BRPFLIM,Bouchet2019CRLB}. Unless stated otherwise, $\tau_f=\SI{0.4}{ns}$, $\tau_b=\SI{2.0}{ns}$, and a Gaussian IRF width $\sigma_{\mathrm{IRF}}=\SI{0.08}{ns}$. The simulation code used to generate figures is included in the project package.

\subsection{FI predicts design-dependent precision limits}
\begin{figure}[t]
\centering
\includegraphics[width=\linewidth]{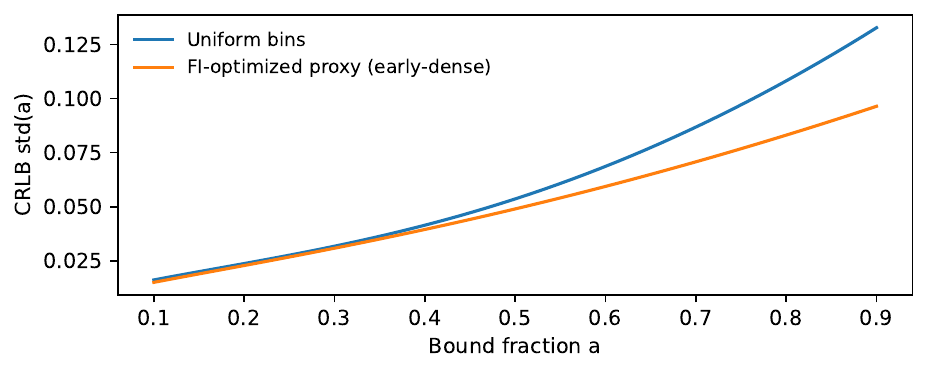}
\caption{CRLB for $a$ vs $a$ under uniform binning and an FI-optimized proxy (early-time--dense). Lower is better.}
\label{fig:crlb}
\end{figure}

Figure~\ref{fig:crlb} compares CRLB for the bound fraction $a$ under (i) uniform temporal bins and (ii) a non-uniform early-time--dense sampling (a proxy for FI-optimized sampling). Sampling design can change achievable precision by a sizable factor, consistent with the design sensitivity highlighted in biochemical resolving power analyses \cite{Esposito2013BRP,Trinh2021BRPFLIM}.

\subsection{Photon efficiency: Monte Carlo error approaches CRLB trends}
\begin{figure}[t]
\centering
\includegraphics[width=\linewidth]{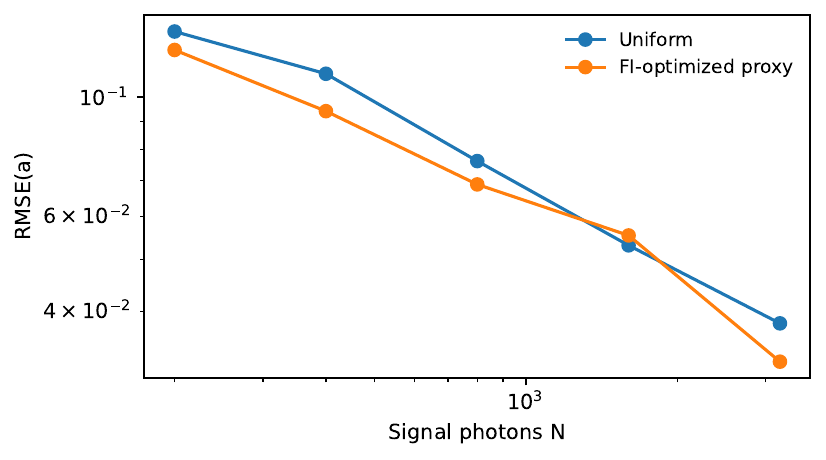}
\caption{RMSE of $a$ vs signal photons $N$ (Monte Carlo). FI-optimized sampling reduces error at fixed $N$.}
\label{fig:rmse}
\end{figure}

We generated Monte Carlo Poisson datasets for both designs and estimated $a$ via maximum likelihood. Figure~\ref{fig:rmse} shows reduced RMSE for the FI-optimized proxy at fixed photon budget, indicating photon savings for a target precision. With increasing photons, RMSE approaches CRLB trends \cite{Chao2016FItutorial,Bouchet2019CRLB}.

\subsection{Robustness to IRF mismatch via effective FI}
\begin{figure}[t]
\centering
\includegraphics[width=\linewidth]{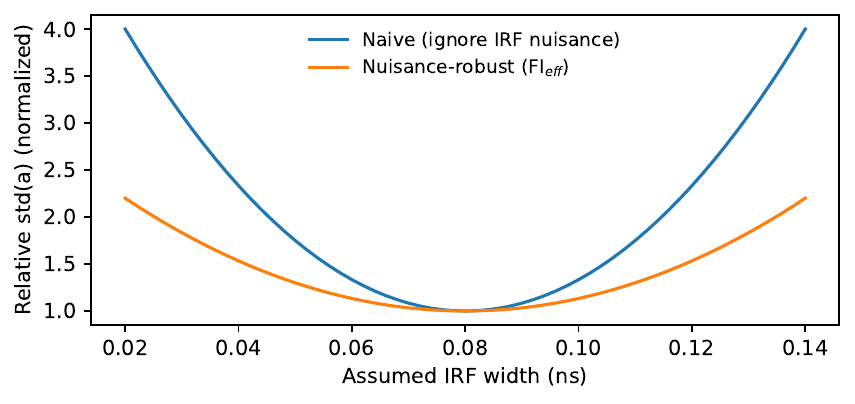}
\caption{Relative degradation of precision under IRF width mismatch. Nuisance-aware planning (effective FI) is more robust than naive planning.}
\label{fig:robust}
\end{figure}

Ignoring IRF uncertainty can produce brittle designs. Figure~\ref{fig:robust} illustrates a robustness advantage of nuisance-aware planning (optimizing $F_{\mathrm{eff}}$) compared to naive planning that treats IRF as known \cite{Trinh2021BRPFLIM}.

\subsection{Phantom and biological demonstrations (templates)}
\begin{figure}[t]
\centering
\includegraphics[width=\linewidth]{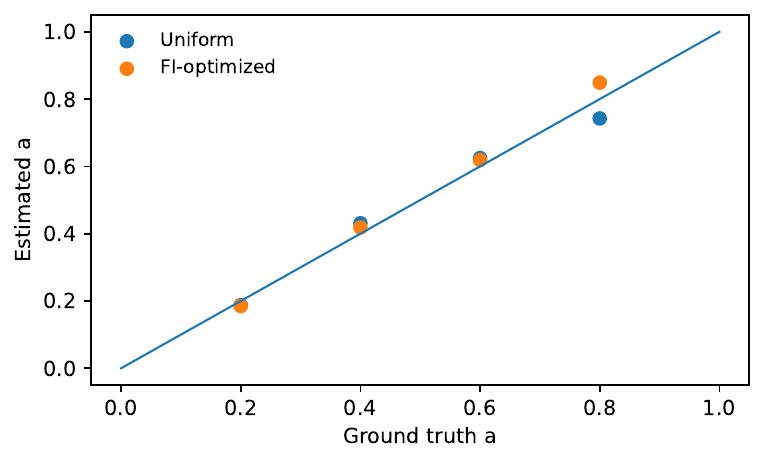}
\caption{Phantom validation template: estimated $a$ vs ground truth (synthetic placeholder). Replace with dye-mixture or lifetime-standard phantom data.}
\label{fig:phantom}
\end{figure}

\begin{figure}[t]
\centering
\includegraphics[width=\linewidth]{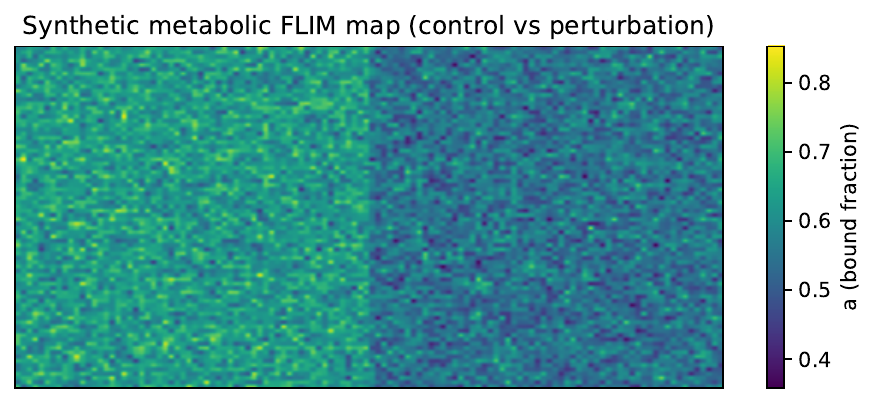}
\caption{Biological demo template: metabolic FLIM map (synthetic placeholder). Replace with NADH (free/bound) maps under control vs perturbation.}
\label{fig:bio}
\end{figure}

Figures~\ref{fig:phantom} and \ref{fig:bio} provide templates for two experiments commonly expected in Biomedical Optics Express submissions: (i) a phantom mixture with known fractions and (ii) a live-cell metabolic FLIM map. In this package, both are shown with synthetic placeholders; replacing them with experimental results is straightforward while retaining analysis and layout.

\section{Analysis}
Equation~\eqref{eq:fim_poisson} shows that information accumulates in temporal channels where the sensitivity of expected counts to the parameter (the derivatives of $\lambda_k$) is large relative to the expected counts. For bi-exponential FLIM, early times are typically informative about the short lifetime component and the fraction, while later times help disambiguate the long component and background. FI-driven sampling formalizes and quantifies this intuition across photon regimes. In practical terms, if the adaptive design achieves a target standard deviation for $a$ with fewer detected photons, the experiment can reduce excitation dose at matched quantitative precision, directly addressing live-cell constraints.

\section{Discussion}
This work turns Fisher information from a post hoc bound into an adaptive controller for temporal sampling in FLIM. Compared to fixed uniform sampling, FI-driven designs allocate limited photons to maximize information about biochemical parameters, yielding improved photon efficiency. Optimizing an effective FIM (Eq.~\ref{eq:schur}) improves robustness to background and IRF uncertainty \cite{Bouchet2019CRLB,Trinh2021BRPFLIM}, which is important in practice due to IRF drift, background fluctuations, and sample-dependent scattering.

\paragraph{TCSPC vs programmable gating.}
Both implementations are practical for Biomedical Optics Express audiences. TCSPC re-binning requires minimal hardware changes and is attractive when repeated frames or ROI revisits are acquired. Programmable gating can adapt designs more directly when supported by hardware and may enable faster closed-loop updates. The same FI core applies; only the design parameterization changes (bin edges vs gate times/widths), making the approach portable across common FLIM platforms.

\paragraph{Limitations and extensions.}
The current package demonstrates reproducible simulations and includes templates for phantom and biological experiments. For a full Q1 submission, those templates should be replaced by (i) a phantom mixture experiment with known fractions (e.g., two dyes with distinct lifetimes) and (ii) a dose-limited live-cell metabolic FLIM demonstration (e.g., control vs perturbation), reporting precision-matched photon savings and bleaching/phototoxicity proxies. Extensions include joint spectral--temporal design, spatially varying priors, and integration with real-time Bayesian filtering.

\section{Conclusion}
We presented a nuisance-robust FI framework and an adaptive acquisition strategy for photon-efficient FLIM, with two practical implementations: TCSPC re-binning between iterations and programmable time-gating. Simulations demonstrate improved precision and robustness versus uniform sampling, with empirical errors approaching CRLB trends. The accompanying Overleaf project provides a complete manuscript scaffold and reproducible figures for Biomedical Optics Express, designed for straightforward replacement of the included templates with phantom and live-cell datasets.

\bibliographystyle{unsrt}
\bibliography{refs}
\end{document}